\documentclass[12pt]{article}
\usepackage{fullpage}
\usepackage{graphicx}

\def\be{\begin{equation}}
\def\ee{\end{equation}}
\def\beq{\begin{eqnarray}}
\def\eeq{\end{eqnarray}}
\def\Gev{{\rm GeV}}
\def\Tev{{\rm TeV}}

\def \lsim{\mathrel{\vcenter
      {\hbox{$<$}\nointerlineskip\hbox{$\sim$}}}}
\def \gsim{\mathrel{\vcenter
      {\hbox{$>$}\nointerlineskip\hbox{$\sim$}}}}

\begin{document}
\begin{flushright}
TIFR/TH/05-13
\end{flushright}
\bigskip
\begin{center}
{\Large\bf Looking for a Heavy Higgsino LSP in Collider and Dark Matter
Experiments} \\[1cm]
Utpal Chattopadhyay$^a$, Debajyoti Choudhury$^b$, Manuel Drees$^c$, Partha
Konar$^d$ and D.P. Roy$^d$
\end{center}
\hspace*{.5cm}
\begin{enumerate}
\item[{$^a$}] Department of Theoretical Physics, Indian Association for the
Cultivation of Science, Raja S.C. Mallik Road, Kolkata 700 032, India
\item[{$^b$}] Department of Physics and Astronomy, University of Delhi, Delhi
  110007, India 
\item[{$^c$}] Physikalisches Institut der Universit\"at Bonn, Nussallee 12,
  53115 
Bonn, Germany
\item[{$^d$}] Department of Theoretical Physics, Tata Institute of Fundamental
Research, Homi Bhabha Road, Mumbai 400 005, India
\end{enumerate}
\bigskip
\bigskip
\begin{center}
\underbar{\bf Abstract}
\end{center}
\bigskip

A large part of the mSUGRA parameter space satisfying the WMAP
constraint on the dark matter relic density corresponds to a higgsino
LSP of mass $\simeq 1$ TeV.  We find a promising signal for this LSP
at CLIC, particularly with polarized electron and positron beams.
One also expects a viable monochromatic $\gamma$-ray signal from its pair
annihilation at the galactic center at least for cuspy DM halo
profiles.  All these results hold equally for the higgsino LSP of
other SUSY models like the non-universal scalar or gaugino mass models
and the so--called inverted hierarchy and more minimal supersymmetry
models.

\newpage

\section*{1. Introduction}
\medskip

Weak scale supersymmetry (SUSY) is the most popular extension of
the standard model (SM) because it is endowed with three unique
features \cite{1}.  It provides (1) a natural solution to the hierarchy
problem of the SM, (2) a plausible candidate for the cold dark matter
of the universe in the form of the lightest superparticle (LSP), and
(3) unification of the SM gauge couplings at the GUT scale.  However
it also suffers from two problems.
\bigskip

\noindent (i) {\it Little Hierarchy Problem:} The LEP limit on the
mass of an SM--like Higgs boson \cite{2},
\be
m_h > 114 \ \Gev,
\label{one}
\ee
requires the average top squark mass to be well above $M_Z$ \cite{3}. In
models where supersymmetry breaking is transmitted to the visible sector at an
energy scale exponentially larger than the weak scale, this implies some
fine-tuning of SUSY parameters to obtain the correct value of $M_Z$.  \bigskip

\noindent {\it (ii) Flavor and CP Violation Problem:} Generic SUSY
models make fairly large contributions to CP violating processes with or
without flavor violation, as represented by the $K$ decay observable
$\epsilon_K$ and the fermion electric dipole moments (EDM) respectively.
Predictions for rates of CP--conserving flavor changing processes, like $\mu
\rightarrow e \gamma$ decays, also often exceed experimental limits. SUSY
parameters have to be chosen carefully to control these contributions. It
should be noted here that the recently advocated split SUSY model \cite{4}
tries to solve the second problem by pushing up the scalar superparticle
masses, but at the cost of dramatically aggravating the first problem.

The minimal supergravity model (mSUGRA) is by far the simplest potentially
realistic model of weak scale supersymmetry \cite{5}. It provides a very
economical parametrisation of superparticle masses and couplings on the one
hand and a natural explanation of the electroweak symmetry breaking (EWSB)
phenomenon on the other \cite{6}. The model is completely specified in terms
of four continuous parameters and one sign, namely
\be
m_0,m_{1/2},A_0,\tan\beta \ {\rm and} \ {\rm sign}(\mu).
\label{two}
\ee
The first three entries represent the universal SUSY breaking scalar and
gaugino masses and trilinear coupling at the GUT scale. $\mu$ is the
supersymmetric Higgs(ino) mass parameter, while $\tan\beta$ is the ratio of
the two Higgs vacuum expectation values. The flavor universality of scalar
soft breaking terms avoids problems with flavor changing processes, while CP
violation will be under control if the parameters in (\ref{two}) are real.

In going down from the GUT scale to the weak scale the $SU_3 \times
SU_2 \times U_1$ gaugino masses evolve like their respective gauge
couplings, i.e.
\beq
\label{three}
M_1 &=& (\alpha_1/\alpha_G) m_{1/2} \simeq (25/60) m_{1/2}, \nonumber
\\[2mm]
M_2 &=& (\alpha_2/\alpha_G) m_{1/2} \simeq (25/30) m_{1/2}, \\[2mm]
M_3 &=& (\alpha_3/\alpha_G) m_{1/2} \simeq (25/9) m_{1/2}. \nonumber
\eeq
The higgsino masses are simply given by the $\mu$ parameter. The scalar
masses at the weak scale are also related by Renormalization Group Equations
(RGE) to the GUT scale mass parameters of eq.(\ref{two}). A very important
scalar mass at the weak scale is the mass of the Higgs boson $H_2$ that
couples to the top quark. This mass appears in the EWSB condition
\be
\mu^2 + M^2_Z/2 = {m^2_{H_1} - m^2_{H_2} \tan^2\beta \over
\tan^2\beta - 1} \simeq -m^2_{H_2}.
\label{four}
\ee
The last equality holds for $\tan\beta \geq 5$, which is favored by
the LEP limit of eq.(\ref{one}) \cite{2,3}. $m^2_{H_2}$ is related to
the GUT scale mass parameters by the solution of its RGE \cite{7},
\be
m^2_{H_2} = C_1 m^2_0 - C_2 m^2_{1/2},
\label{five}
\ee
where we have dropped contributions $\propto A_0$ for simplicity since they do
not play any important role here. The coefficients $C_1,C_2$ depend on the
gauge and Yukawa couplings. Thanks to the large negative contribution from the
top Yukawa coupling, $C_2 \simeq 2$. On the other hand, $|C_1| \ll 1$, its
value and sign depending on the exact values of SM parameters (in particular,
on $m_t$ and $\alpha_s$), on the scale $m_{\rm SUSY}$ where the RG evolution
is terminated, and on $\tan\beta$, with smaller $\tan\beta$ favoring smaller
(possibly negative) $C_1$. This makes it easy to obtain a negative value of
$m^2_{H_2}$ as required by the EWSB condition (\ref{four}). This is the
so-called radiative EWSB mechanism \cite{6}.

Combining eqs. (\ref{four}) and (\ref{five}), one sees that $\mu^2$ is related
to $m^2_0$ and $m^2_{1/2}$ by an ellipsoidal equation if $C_1 < 0$, in
particular for low values of $\tan\beta$ ($< 5$). However, for moderate to
large values of $\tan\beta$ ($> 5$), favored by the LEP limit of
eq.(\ref{one}), and large $m_{\rm SUSY}$, $C_1$ becomes positive, leading
to a hyperbolic equation.  These two cases have been described as ellipsoidal
and hyperbolic branches of mSUGRA \cite{8}.  One sees from eqs.
(\ref{three},\ref{four},\ref{five}) that in the first case $M_1 < |\mu|$; and
the lightest neutralino (LSP) is the Bino ($\tilde B$).  However in the
phenomenologically favored case of moderate to large $\tan\beta (\geq 5)$ one
can have both $M_1 < |\mu|$ and $M_1 > |\mu|$, corresponding to a Bino and a
higgsino LSP respectively \cite{8}.  We shall concentrate in this case.  Since
the sign of $\mu$ is not important for our analysis, we shall choose only
positive sign for simplicity.

In the next section we study the dark matter (DM) relic density in the Bino
and higgsino LSP domains of mSUGRA model following the second paper of
ref.\cite{8}. The relic density has been computed using the MicrOMEGAs code
\cite{9}. A comparison with cosmological data \cite{10} on the relic DM
density shows that a large fraction of mSUGRA points satisfying this data come
from the higgsino LSP domain with $|\mu| \simeq 1$ TeV. In the following two
sections we study the prospects of detecting the higgsino LSP in collider and
DM search experiments, respectively. We find good prospect of detecting this
particle at a 3 TeV linear collider like CLIC. There is also a good prospect
of detecting it in the form of TeV scale gamma ray line from DM
pair-annihilation in the galactic center for favorable profiles of galactic DM
distribution.  In the next section we shall show that all our results hold not
only in mSUGRA but in a host of other SUSY models as well, which can naturally
accommodate a higgsino LSP.  Finally we shall conclude with a summary of our
results.  \bigskip

\section*{2. Higgsino LSP as DM in mSUGRA}
\medskip

\begin{figure}
\begin{center}
\includegraphics[width=8cm]{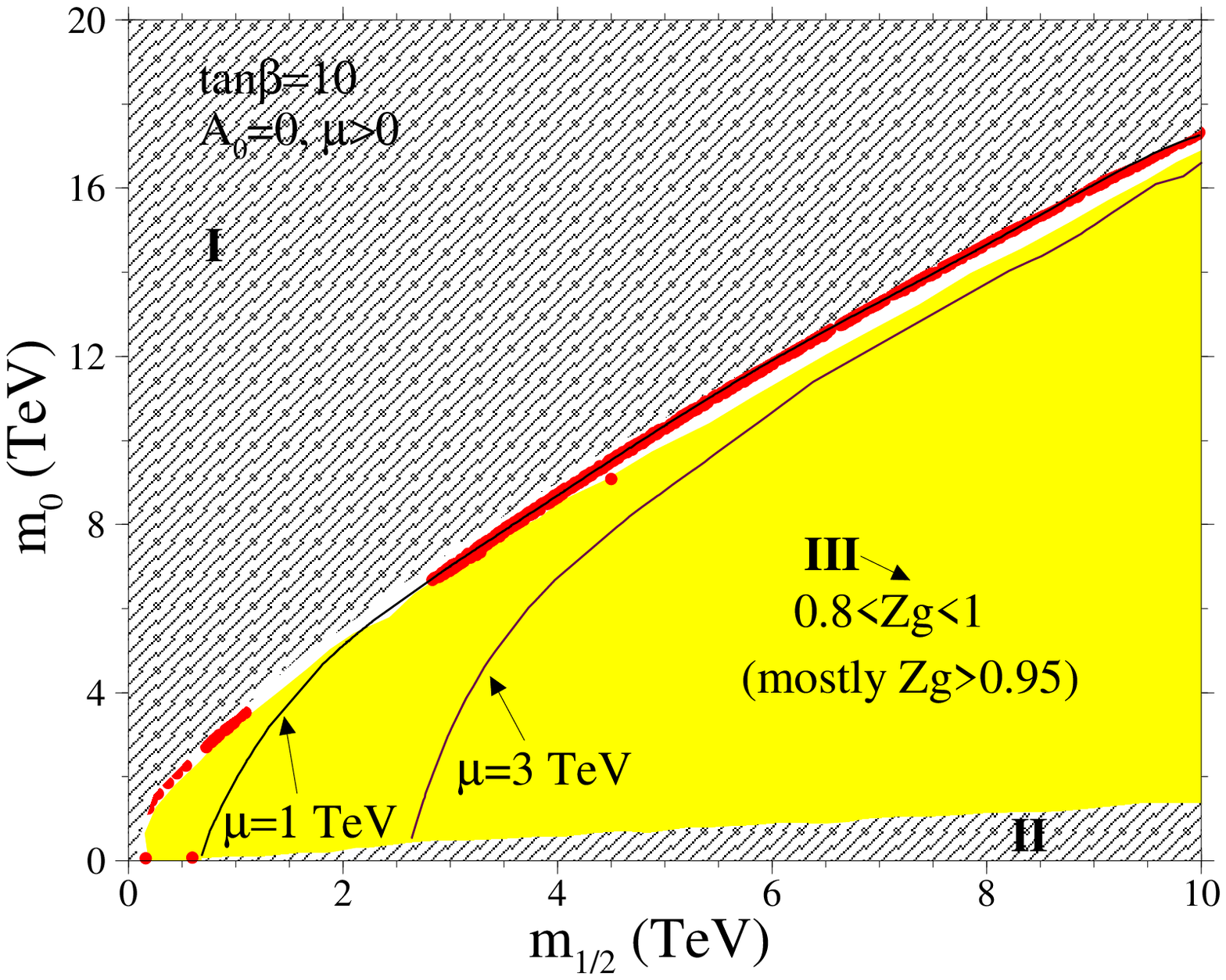}
\includegraphics[width=8cm]{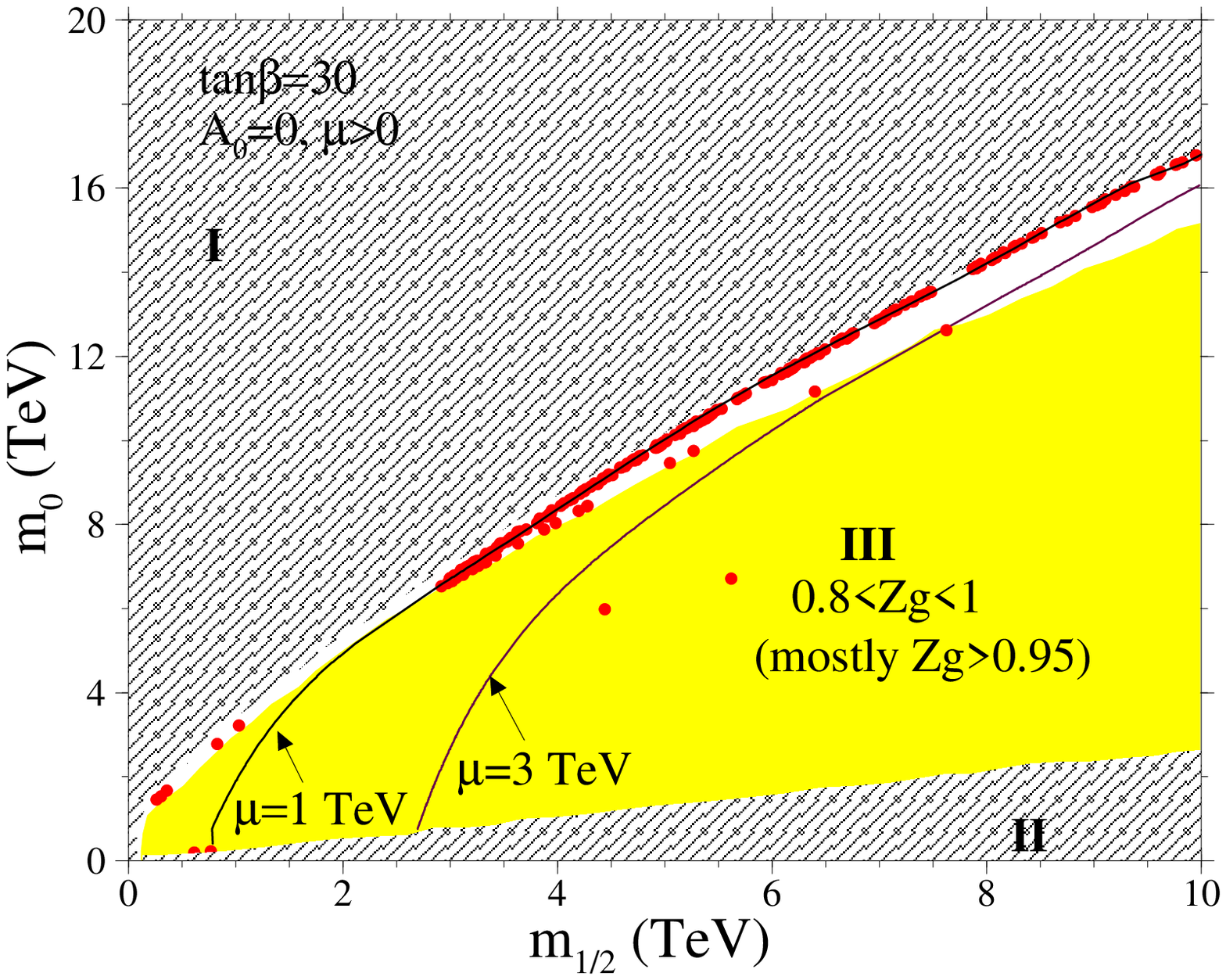}
\end{center}
\caption{mSUGRA parameter space for $m_t = 178$ GeV, $A_0 = 0$ and 
  $\tan\beta=10$ (left) and 30 (right). The patterned regions marked I and II
  are disallowed by the EWSB condition and the constraint of a neutral LSP,
  respectively. In the large yellow region the LSP is Bino--like. Red points
  satisfy the constraint (8) on the dark matter relic density.}
\end{figure}

Figs. 1 show the mSUGRA parameter space satisfying the EWSB condition for a
moderate and a large value of $\tan\beta$. We have used our own code for the
solution of the relevant RGE and the treatment of EWSB, including dominant
loop corrections. The upper edge of the allowed (white) region corresponds to
the hyperbolic boundary from eqs.(\ref{four},\ref{five}) for the LEP limit of
$|\mu| \geq 100$ GeV \cite{2}. In fact, the bulk of this disallowed region (I)
corresponds to $\mu^2 < 0$, i.e. no EWSB. In the bottom strip (II) the tau
slepton $\tilde\tau_1$ becomes the LSP. This is disallowed by the
astrophysical constraint requiring a neutral LSP \cite{2}. Over the allowed
region the LSP is the lightest neutralino, which can be a combination of
gaugino and higgsino states, i.e.
\be
\chi \equiv \tilde\chi^0_1 = N_{11} \tilde B + N_{12} \tilde W_3 +
N_{13} \tilde H_1^0 + N_{14} \tilde H_2^0.
\label{six}
\ee
The gaugino component of the LSP is defined by the fraction
\be
Z_g = N^2_{11} + N^2_{12}.
\label{seven}
\ee
The yellow (light shaded) region in Fig. 1 corresponds to a dominantly
gaugino (in fact Bino $\tilde B$) LSP.

One sees from Fig. 1 that in most of the mSUGRA parameter space the LSP is
dominantly $\tilde B$. Notice, however, that the bulk of this region is
disallowed by the constraint on the DM relic density from cosmological data,
in particular from the WMAP satellite experiment, \cite{10}
\be
\Omega_\chi h^2 = 0.113 \pm 0.017,
\label{eight}
\ee
where $h = 0.71 \pm 004$ is the Hubble constant in units of 100
km/(s$\times$Mpc) \cite{2} and $\Omega$ is the relic density in units of the
critical density. In fact, one usually finds an overabundance of DM relic
density $(\Omega_\chi > 1)$. This is because $\tilde B$ does not carry any
gauge charge and hence does not couple to the gauge bosons. A Bino--like LSP
therefore mostly annihilates via the exchange of sfermions, $\chi\chi
{\buildrel {\tilde f} \over \longrightarrow} f\bar f$, which is suppressed by
the large sfermion mass. Only in some special cases like $m_\chi \simeq
m_{\tilde\tau_1}$ and $m_\chi \simeq m_A/2$ can one get large co--annihilation
$\chi\tilde\tau_1 {\buildrel \tau,\tilde\tau \over \longrightarrow}
\tau\gamma$ and pair annihilation $\chi\chi {\buildrel A \over
  \longrightarrow} b\bar b, t\bar t$ rates respectively \cite{11}.
Correspondingly one can see a few (red) points of acceptable $\tilde B$ DM
density near the lower boundary (co--annihilation); in mSUGRA resonant
$A-$exchange becomes possible only for $\tan \beta \geq 50$.\footnote{There is
  also a small allowed region \cite{hpole} with $m_\chi \simeq m_h/2$, $h$
  being the lighter CP--even Higgs boson; this is however not visible at the
  scales chosen in Fig.~1.}

Most of the points satisfying the DM relic density constraint (\ref{eight})
are seen to lie very near the hyperbolic boundary \cite{9}. The few points
near the lower end of this boundary correspond to the so--called focus point
region \cite{12}, where the LSP has a significant higgsino component, although
it may be still dominated by $\tilde B$. Such an LSP couples via its higgsino
component to $W$ and $Z$ bosons, and through gaugino--higgsino mixing to Higgs
bosons, and can thus annihilate into both fermionic and bosonic final states.
LHC signatures for the focus point region have been investigated in \cite{13}.
Note however that the large majority of DM--allowed points lie on the
\be
m_\chi \simeq \mu = 1 \ \Tev
\label{nine}
\ee
contour. For the chosen value of the top mass, $m_t = 178$ GeV, the DM
constraint (\ref{eight}) is satisfied on this contour for $m_{1/2} \geq 3$ TeV
and $m_0 \geq 6.5$ TeV; for the new preliminary world average top mass of 173
GeV \cite{newtop}, the lower bound on $m_0$ would be reduced to $\simeq 5.5$
TeV. This is the higgsino LSP domain of mSUGRA (gaugino fraction $Z_g \lsim
0.1$).\footnote{There must be allowed points also in between the focus point
  and TeV higgsino--LSP regions. However, this DM--allowed strip is very
  narrow, since the transition between a lighter higgsino as LSP, with too
  small a relic density, and Bino--like LSP with much too high a relic density,
  is very rapid. The scan of parameter space used in Figs.~1 therefore found
  no allowed points for $m_0$ between 4 and 6.5 TeV.} In this region there is a
near degeneracy among the lighter chargino and neutralino states, i.e.
\be 
m_{\chi} \simeq m_{\tilde \chi^0_2} \simeq m_{\tilde \chi^+_1} \simeq \mu
\simeq 1 \ \Tev.  
\label{ten}
\ee
So the major annihilation processes correspond to the pair and
co--annihilation \cite{14}
\be
\tilde \chi^0_i \tilde \chi^0_i \ {\buildrel \tilde \chi^+_1 (\tilde \chi^0_j)
  \over \longrightarrow} \ W W(ZZ), \ 
\tilde \chi^0_i \tilde \chi^+_1 \ {\buildrel W \over \longrightarrow} \ \bar
f_1 f_2, \ 
\tilde \chi^0_1 \tilde \chi^0_2 \ {\buildrel Z \over\longrightarrow} \ \bar f
f, 
\label{eleven}
\ee
where $i = 1,2$ and $j= 2 \ (1)$ for $i = 1 \ (2)$. Although $Z$ couples to a
pair of $\chi$ via their higgsino components, the coupling is proportional to
the difference $N^2_{13} - N^2_{14}$ \cite{17}. Hence it vanishes in the limit
of $M_1, |\mu| \gg M_Z$, where the $\tilde \chi^0_{1,2}$ eigenstates
correspond to the symmetric and antisymmetric combinations of $\tilde H_1^0$
and $\tilde H_2^0$. In the same limit, the off--diagonal $Z \tilde \chi_1^0
\tilde \chi_2^0$ coupling reaches its maximal value. Thanks to the
annihilation processes (\ref{eleven}), the string of points satisfying the
constraint (\ref{eight}) continues indefinitely upwards on the $\mu = 1$ TeV
contour, whereas all other DM--allowed regions in mSUGRA are finite in the
$(m_{1/2}, m_0)$ plane. In this sense the constraint (\ref{eight}) favors the
higgsino LSP domain of the mSUGRA model.

On the other hand, Figs.~1 imply that in this domain all superparticles are
quite heavy. We just saw that even the LSP has a mass near 1 TeV. Moreover, we
needed $m_{1/2} \geq 3$ TeV and $m_0 \geq 6.5$ TeV for $m_t = 178$ GeV. This
means that the electroweak gaugino $(\tilde B,\tilde W)$ masses are at least
in the few TeV range, while the masses of gluinos and all scalars (except for
the lightest Higgs boson) are near 10 TeV or even higher. This aggravates the
``little hierarchy'' problem considerably; however, the finetuning required is
still very much smaller than in split supersymmetry \cite{4}.

Higgsino and gaugino masses in the above range are still compatible with gauge
coupling unification within the uncertainty of GUT scale thresholds. A
sfermion mass scale near 10 TeV is adequate to solve the problems of flavor
and CP violation even without assuming flavor universality \cite{15}. This
leads us to more general SUSY models, which we will comment on in Sec.~5. In
the next two sections we investigate the prospects of probing scenarios with
heavy higgsino--like LSP in collider and dark matter experiments. These
phenomenological investigations are largely model--independent, so long as the
remaining sparticles lie significantly above the higgsino--like states.

\section*{3. Probing the Higgsino LSP Region in Collider Experiments}

Sfermion and gluino masses of $\gsim 10$ TeV and electroweak gaugino masses of
at least a few TeV put them well out of reach of the LHC. The only
superparticles which can be produced there with significant rates are the
nearly degenerate charged and neutral higgsinos, $\chi^+_1$ and $\chi^0_{1,2}$
of mass $\simeq 1$ TeV. We have computed the mass differences including
radiative corrections \cite{17} and found them to be restricted to the range
\be 
\delta m_c =
m_{\chi^+_1} - m_{\chi^0_1} < 10 \ \Gev,
\label{sixteen}
\ee
with $m_{\chi^0_2} - m_{\chi^0_1} \simeq 2\delta m_c$ for $\tan\beta \gg 1$.
Thus the $\tilde \chi_1^\pm$ and $\tilde \chi_2^0$ decay products will be too
soft to be detected efficiently on top of the underlying event at a hadron
collider. Therefore one has to tag the pair production of higgsinos at the
LHC. The by far best tag is provided by the two forward jets $j$ in $\tilde
\chi$ pair production via vector boson fusion \cite{18},
\be
pp \rightarrow \chi^\pm_1 \chi^0_i jj, \chi^+_1 \chi^-_1 jj, \chi^0_1 \chi^0_2
jj \ \ (i=1,2). 
\label{eighteen}
\ee
We have computed the resulting higgsino signal at the LHC closely following
\cite{18} and a similar investigation for an invisibly decaying Higgs signal
in \cite{19}. The selection criteria used are: (i) two forward jets in
opposite hemispheres with $E^j_T > 40$ GeV and $2 < |\eta_j| < 5$; (ii)
$\Delta \eta_{jj} > 4$; (iii) $M_{\rm inv} (jj) > 1200$ GeV; (iv)
${E\!\!\!/}_T > 100$ GeV; (v) $\Delta \phi_{jj} < 57^\circ$ and (vi) the
central jet veto as defined in ref. \cite{19}. The backgrounds come from $Z
(\rightarrow \nu\nu)$ and $W(\rightarrow \ell\nu)$ production via electroweak
(vector boson fusion) and QCD (higher order Drell--Yan) processes where $\ell$
is assumed to escape detection for $p^\ell_T < 10$ GeV. Following \cite{19} we
have assumed the efficiency of the central jet veto to be 0.9, 0.82 and 0.28
for the signal, electroweak and QCD backgrounds respectively. The total
background is 64 fb assuming conservatively the renormalization scale for
$\alpha_s$ to be the lower jet-$E_T$. One expects to measure the background
to a high precision of $\sim 1.2\%$ from the visible $Z \rightarrow \ell
\bar\ell$ and $W \rightarrow \ell \nu$ events \cite{19}. Adding this
uncertainty in quadrature to the statistical error on the background, and
assuming an integrated luminosity of 100 fb$^{-1}$, one thus needs a signal
cross section of at least 5.5 fb for a 5$\sigma$ discovery. Unfortunately the
cross section after cuts for the production of higgsino--like charginos and
neutralinos with mass near 1 TeV is several orders of magnitude below this
value. We therefore conclude that the LHC will not be able to probe the region
of parameter space we are interested in.

The most promising machine for detecting a 1 TeV higgsino LSP is the proposed
3 TeV linear $e^+e^-$ collider CLIC \cite{20}. We shall follow the strategy of
ref.\cite{21} for computing the signal and background cross sections. The same
strategy has been followed by the LEP experiments for setting mass limits on a
higgsino LSP \cite{2}; in particular the OPAL collaboration \cite{22} has used
it to set a mass limit of 90 GeV. The higgsino pair production is tagged by a
photon from initial--state radiation (ISR), i.e.\footnote{The cross sections
  for $\tilde \chi_i^0 \ (i=1,2)$ pair production are negligible, since the
  diagonal $\tilde \chi_i^0 \tilde \chi_i^0 Z$ couplings are very small, as
  remarked earlier.}
\be
e^+e^- \rightarrow \gamma \chi^+_1 \chi^-_1, \gamma \chi^0_1 \chi^0_2 \, .
\label{twenty}
\ee
If the $\tilde \chi_1^\pm$ and $\tilde \chi_2^0$ decay products remain
undetected, the main physics background is
\be
e^+e^- \rightarrow \gamma \nu \bar\nu.
\label{twentyone}
\ee
The photon is required to have an angle $\theta > 10^\circ$ relative
the beam axis. Moreover it is required to satisfy
\be
E^\gamma_T > E^{\gamma \ {\rm min}}_T = \sqrt{s} {\sin\theta_{\rm min}
\over 1 + \sin\theta_{\rm min}},
\label{twentytwo}
\ee
which vetos the radiative Bhabha background $e^+e^- \rightarrow \gamma
e^+e^-$, by kinematically forcing one of the energetic $e^\pm$ to
emerge at an angle $> \theta_{\rm min}$.  At CLIC energy of $\sqrt{s}
= 3$ TeV,
\be
E^{\gamma \ {\rm min}}_T = 50 \ (100) \ \Gev \ {\rm for} \ \theta_{\rm
min} = 1^\circ (2^\circ).
\label{twentythree}
\ee
The OPAL detector has instrumentation for $e^\pm$ detection down to
$\theta_{\rm min} = 2^\circ$, while it seems feasible to have it down to
$1^\circ$ at the future linear colliders \cite{21}. We shall show results for
both $E^{\gamma \ {\rm min}}_T = 50$ and 100 GeV. We shall also impose the
recoil mass cut
\be
M_{\rm rec} = \sqrt{s} (1 - 2E^\gamma/\sqrt{s})^{1/2} > 2m_\chi,
\label{twentyfour}
\ee
which is automatically satisfied by the signal (\ref{twenty}). 
Fake photon background processes have been effectively suppressed by
the OPAL collaboration \cite{22} by requiring photon isolation and a
minimum value for the total $p_T$, which are automatically satisfied
by the signal as well as the $e^+e^- \rightarrow \gamma \nu \bar\nu$
background. Therefore we shall not impose these requirements.

\begin{figure}[h!]
\begin{center}
\includegraphics[width=16cm]{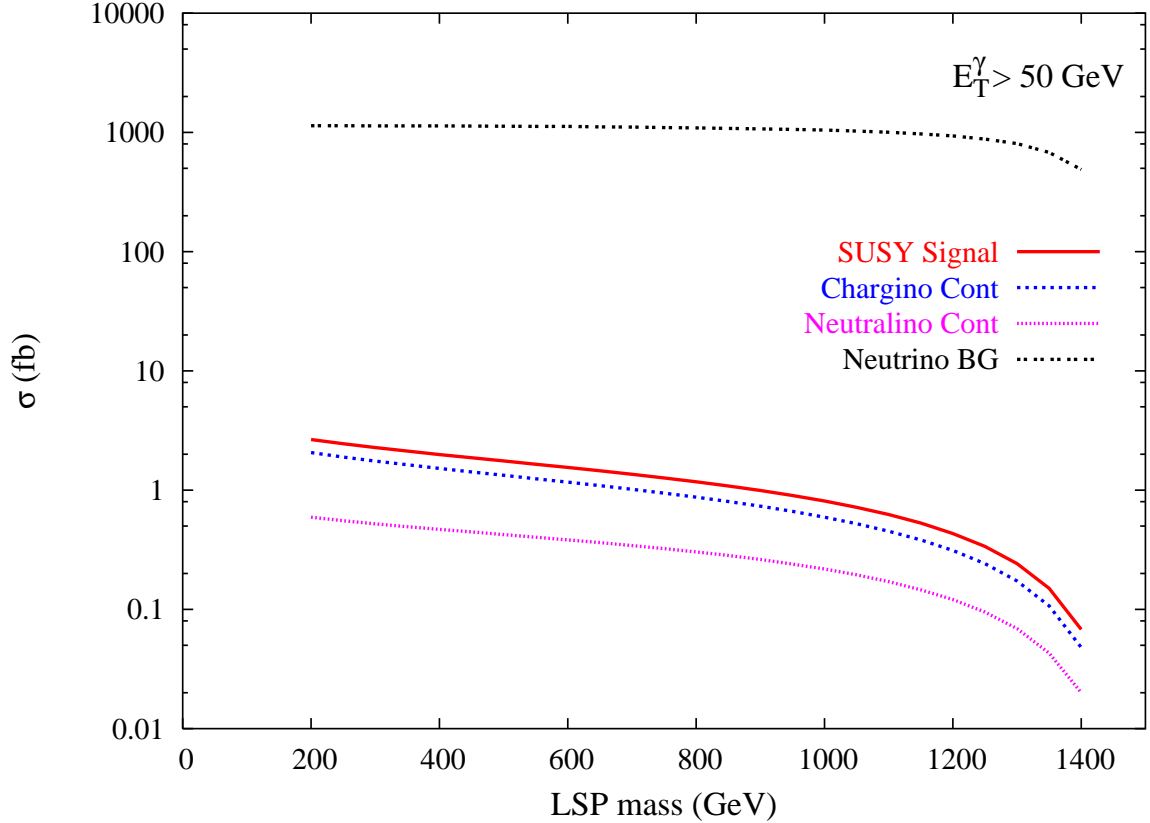}
\end{center}
\caption{Cross sections for the higgsino signal
  (\ref{twenty}) and the neutrino background (\ref{twentyone}) at CLIC
  $(\sqrt{s} = 3)$ TeV, produced with a photon tag of $E^\gamma_T > 50$ GeV.
  Initial state radiation is included.}
\end{figure}

Fig.~2 shows the signal and the background cross sections from (\ref{twenty})
and (\ref{twentyone}) respectively against the higgsino LSP mass, for
$E^{\gamma {\rm min}}_T = 50$ GeV. In calculating these cross sections, we
have included initial state radiation (ISR) effects by convoluting the hard $2
\rightarrow 3$ cross sections with electron distribution functions, as
described in ref.\cite{dg1}. This allows background events with on--shell $Z$
boson, if there is an energetic ISR photon going down one of the beam pipes;
ISR therefore increases the total background by a few \%. We also computed the
higher--order background process
\be \label{bnew}
e^+e^- \rightarrow Z \nu \bar\nu \gamma, \ {\rm where} \ Z \rightarrow \nu
\bar\nu,
\ee
and found it to contribute about 10 fb after cuts -- much less than the
background (\ref{twentyone}), but still significantly more than the signal.

The signal cross section is reduced by ISR by $\sim 10\%$ for $m_\chi = 1$
TeV, since it effectively reduces the amount of phase space available. The
same effect increases the signal for smaller LSP mass, since it increases the
$s-$channel photon and $Z$ propagators. The signal is dominated by the
chargino pair production. For 1 TeV higgsino mass one expects a signal cross
section of only $\sim 0.8$ fb against a background of $\sim 1050$ fb.  Thus
for the projected CLIC luminosity of $1000 \ {\rm fb}^{-1}$ one expects 800
signal events against a background of $10^6$, corresponding to $N_S/\sqrt{N_B}
\simeq 0.8$ only. Evidently it is a hopeless situation unless one can suppress
the background (\ref{twentyone}) by identifying the soft $\chi^\pm_1$ and
$\tilde \chi_2^0$ decay products. This remains true when the cut on
$E^\gamma_T$ is increased to 100 GeV (not shown).

The method of identifying these particles would depend on the decay length
$c\tau$, which depends strongly on the mass difference $\delta m_c$. This
decay length has been estimated in ref. \cite{21} for a specific model of an
iso--triplet chargino. It is shown there that for $\delta m_c \leq 1$ GeV one
expects to detect the chargino track and/or a decay $\pi^\pm (\ell^\pm)$ track
with displaced vertex in a standard micro--vertex detector. One can easily
check that the decay length of the charged higgsino is about twice as large as
the iso--triplet chargino of \cite{21}. Hence these tracks should be even
more clearly detectable in this case.

But one expects prompt chargino decay for $\delta m_c > 1$ GeV, which holds
over most of our parameter space of interest. For this case the OPAL
collaboration \cite{22} has found that the resulting charged tracks can be
detected with $\geq 50\%$ efficiency for the signal, and used it to eliminate
the $\gamma \nu \bar\nu$ background. For the present case such an efficiency
corresponds to a respectable signal size of $\gsim 400$ events. However, a new
problem arises at future linear colliders, which did not occur at LEP. The
large charge density in the bunches gives rise to ``beamstrahlung'' when the
two bunches cross. The collision of beamstrahlung photons can then form an
underlying event containing several soft particles \cite{dg}. If this happens
in the same bunch crossing as a hard $e^+e^- \rightarrow \gamma \nu \bar\nu$
annihilation, one obtains a similar final state topology as in the signal.

It is not possible to speculate at this stage on the level of this underlying
event background at CLIC. All we can say is that an underlying event
resembling the $\tilde \chi_2^0$ or $\tilde \chi_1^\pm$ decay products must
not occur in more than 1\% of all bunch crossings. Neglecting efficiencies,
this would correspond to $\sim 10^4$ background against $\sim 800$ signal
events for $m_\chi = 1$ TeV -- i.e.  $N_S/\sqrt{N_B} \sim 8$. We will see
below that one can tolerate a higher level of underlying events from
beamstrahlung if the $e^+e^-$ beams are polarized. Finally, we note that
beamstrahlung will also change the effective $e^\pm$ beam spectra, and hence
the cross sections (\ref{twenty}) and (\ref{twentyone}). These effects will
need to be included once the beam characteristics have been fixed.

The reason for the large cross section for the background (\ref{twentyone})
compared to the signal processes (\ref{twenty}) is the $t-$channel $W$
exchange contribution to the background. This can be suppressed with right--
(left--)handed polarization of the $e^- \ (e^+)$ beam. In fact it is easy to
see that for 100\% polarization of one of the beams the background cross
section will go down to the level of the signal. This is not feasible, of
course. What we shall do instead is to estimate the signal and background for
the same beam polarizations as envisaged for the ILC \cite{23}, i.e.
\be
P_{e^-} = 0.8 \ {\rm (mostly \ right-handed) \ and} \ P_{e^+} = -0.6 \ {\rm
  (mostly \ left-handed)} .
\label{twentyfive}
\ee
It is easy to check that this corresponds to the following fractional
luminosities
\be
e^-_R e^+_L : e^-_L e^+_R : e^-_L e^+_L : e^-_R e^+_R = 0.72 : 0.02 :
0.08 : 0.18,
\label{twentysix}
\ee
while each was 0.25 in the unpolarized case. The dominant contribution to the
background (\ref{twentyone}) from $t$-channel $W$ exchange contributes only to
the second combination $e^-_L e^+_R$.  Hence it is suppressed by a factor of
$.02/.25 = .08$. The higher--order background (\ref{bnew}) is suppressed by a
similar factor. One can also check that the $\tilde\chi^+_1 \tilde\chi^-_1$
and $\tilde\chi^0_1 \tilde\chi^0_2$ contributions to the signal (\ref{twenty})
are modified by factors of 0.6 and 1.3 respectively, resulting in an overall
suppression of the total signal by a factor 0.8. Fig.~3 shows the total signal
and background cross sections for $E^\gamma_T > 50$ and 100 GeV. In either
case one gets a $N_S/\sqrt{N_B} \simeq 2.1$ for the CLIC luminosity of 1000
fb$^{-1}$.  But one has a better $N_S/N_B \simeq 400/39,000$ events for the
$E^\gamma_T > 100$ GeV cut. Recall that this cut requires instrumentation down
to $2^\circ$ instead of $1^\circ$ to eliminate the $\gamma e^+e^-$ background.
Hence this harder cut seems advantageous to us.

\begin{figure}[h!]
\begin{center}
\includegraphics[width=16cm]{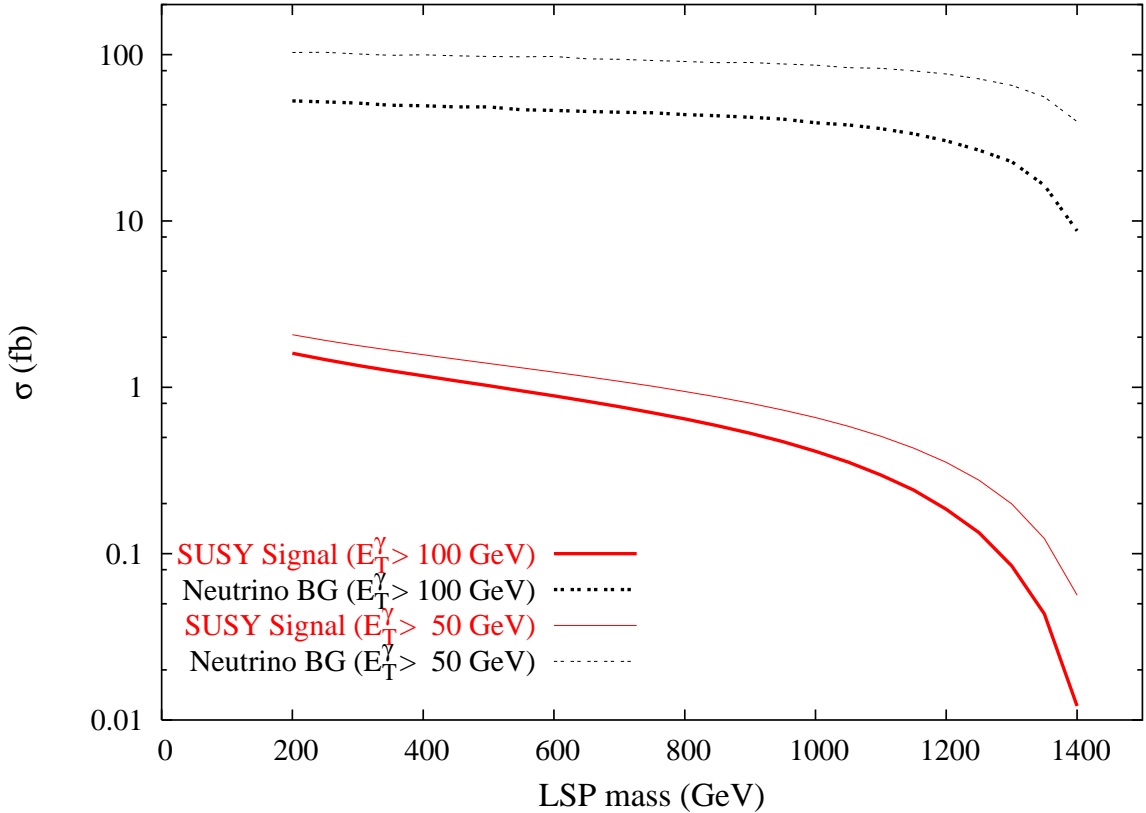}
\end{center}
\caption{Cross sections of the higgsino signal (\ref{twenty}) and neutrino
  background (\ref{twentyone}) at CLIC with polarized $e^-$ and $e^+$
  beams. Initial state radiation is included.}
\end{figure}

Fig. 4 shows the recoil mass distribution of the background (\ref{twentyone})
along with that of a 1 TeV higgsino signal (\ref{twenty}) for both
$E^\gamma_T$ cuts and polarized beams; ISR has again been included. As
discussed above the background can be suppressed if the soft $\tilde
\chi_1^\pm$ or $\tilde \chi_2^0$ decay products can be detected. In case of
the background, such soft particles can only come from beamstrahlung reactions
like $\gamma\gamma \rightarrow \pi^+ \pi^-,\ell^+\ell^-,\cdots$ underlying the
background (\ref{twentyone}). Not all such reactions will lead to events with
similar characteristics as the signal. For example, one can envision applying
cuts on the angular distribution of the soft particles, which tend to peak at
small angles in two--photon events, but are quite central for most signal
events. Another possible discriminator is the $p_T$ imbalance of the soft
particles (i.e., not counting the hard tagging photon), which is expected to
be larger for the signal than for the background. In devising such cuts, the
characteristics of the (largely non--perturbative) background can be taken
from measurements in the pure background region $M_{\rm rec} < 2 m_\chi$.
Comparing the observed cross section for the remaining background over the
$M_{\rm rec} < 2$ TeV region with the prediction of Fig.~4 would give an
estimate of the fraction of surviving background due to beamstrahlung. Since
this fraction should be independent of $M_{\rm rec}$, one can use this to
estimate the cross section of the surviving background in the $M_{\rm rec} >
2$ TeV signal region. Any excess over this estimate would represent the
higgsino signal. One might even be able to estimate the higgsino mass from the
threshold of the excess cross section. It is easy to see that underlying
events from beamstrahlung at the 10\% level (after cuts) correspond to a
reduction of the background to 10\% and hence will increase $N_S/\sqrt{N_B}$
ratio from 2 to $\sim 6$. Thus with polarized beams one can tolerate the
underlying event at the 10\% level.

\begin{figure}[h!]
\begin{center}
\includegraphics[width=16cm]{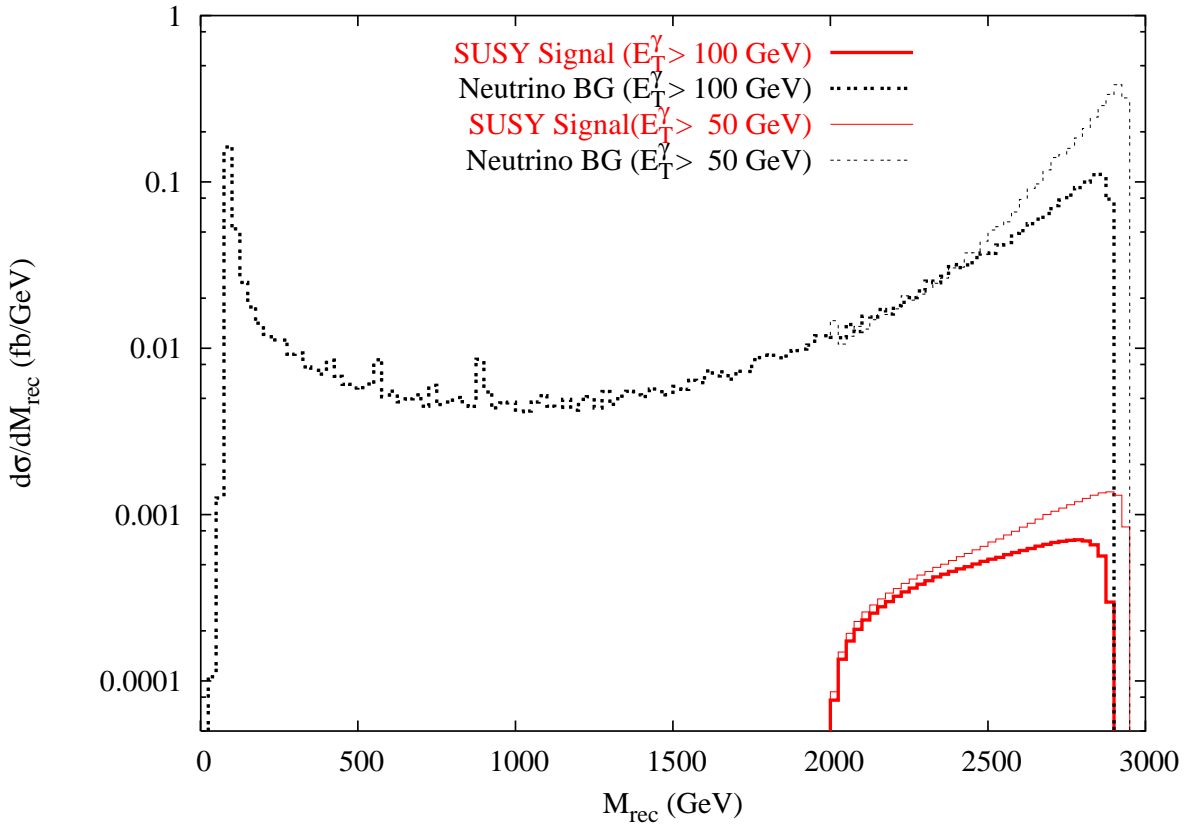}
\end{center}
\caption{The recoil mass distributions of a 1 TeV higgsino signal
  (\ref{twenty}) and the neutrino background (\ref{twentyone}) at CLIC with
  polarized $e^-$ and $e^+$ beams. Initial state radiation is included.} 
\end{figure}

\section*{4. Higgsino LSP Search in DM experiments}

One can see from the second paper of ref. \cite{8} that the higgsino LSP
signal is too small to be measurable in direct dark matter search experiments.
The reason is that the signal comes from spin--independent $\chi p$
scattering, which is dominated by the Higgs boson $(h, \, H)$ exchange. Since
its coupling to a $\chi$ pair is proportional to the product of their higgsino
and gaugino components, it is very small for a higgsino dominated LSP.  The
signal is further suppressed by the large LSP mass.

We have also checked that the neutrino signal coming from the $\chi$ pair
annihilation at the solar core is too small to be measurable at an IceCUBE
size detector. Here the signal size is determined by the spin--dependent $\chi
p$ scattering cross section via $Z$ boson exchange, which is very small due to
the suppressed diagonal $Z \chi \chi$ coupling; see the remark following
eq.(\ref{eleven}).

The most promising signal for TeV higgsino DM comes from the pair annihilation
processes 
\be
\chi \chi \rightarrow \gamma \gamma, \chi\chi \rightarrow \gamma Z,
\label{twentyseven}
\ee
resulting in a monochromatic $\gamma-$ray line \cite{24,25}. The dominant
contributions to these processes come from $W^\pm \chi^\mp_1$
loops, and are suppressed by only a $M^2_W$ factor in the denominator
instead of $m^2_\chi$. This results in a large cross section for
$\gamma-$ray production from (\ref{twentyseven}) for a TeV scale
higgsino,
\be
v \sigma_{\gamma\gamma} \sim v \sigma_{\gamma Z} \sim 10^{-28} \ {\rm
cm}^3 {\rm s}^{-1},
\label{twentyeight}
\ee
where $v$ is the velocity of the DM particles in their cms frame. The
resulting gamma ray flux\footnote{It has been pointed out very recently
  \cite{newan} that tree--level higher order processes, in particular $\chi
  \chi \rightarrow W^+ W^- \gamma$, can increase the flux of photons with
  $E_\gamma \simeq m_\chi$ by up to a factor of 2. While significant, this
  enhancement is still much smaller than the uncertainty coming from the DM
  distribution near the center of the galaxy, as discussed below.} coming from
an angle $\psi$ relative to the galactic center is given by
\be
\phi_\gamma (\psi) = {N_\gamma v\sigma \over 4\pi m^2_\chi} \int_{\rm
line \ of \ sight} \rho^2(\ell) d\ell(\psi),
\label{twentynine}
\ee
where $\rho(\ell)$ is the dark matter energy density and $N_\gamma =
2$ (1) for the $\gamma\gamma \ (\gamma Z)$ production process. This can
be rewritten as \cite{24}
\be
\phi_\gamma (\psi) = 1.87 \times 10^{-14} \left({N_\gamma v \sigma
\over 10^{-28} \ {\rm cm}^3 {\rm s}^{-1}}\right) \left({1 \ \Tev \over
m_\chi}\right)^2 J(\psi) {\rm cm}^{-2} {\rm s}^{-1} {\rm sr}^{-1} \, ,
\label{thirty}
\ee
where
\be
J(\psi) = \int_{\rm line \ of \ sight} \rho^2(\ell) d\ell(\psi)/ \left[ (0.3 \
\Gev/{\rm cm}^3)^2 \cdot 8.5 \ {\rm kpc} \right]
\label{thirtyone}
\ee
is the line integral scaled by the squared DM mass density in our neighborhood
and by our distance from the galactic center.

Several Atmospheric Cerenkov Telescopes (ACT) have started recording or are on
their way to record such $\gamma-$rays from the galactic center -- i.e. MAGIC
and VERITAS in the northern hemisphere and H.E.S.S. and CANGAROO in the south.
One generally expects a concentration of DM in the galactic center; but its
magnitude has a large uncertainty depending on the assumed profile of the DM
halo density distribution \cite{26,27,28}. The cuspy NFW profile \cite{26}
corresponds to
\be
\langle J(0)\rangle_{\Delta \Omega = .001} \simeq 1000,
\label{thirtytwo}
\ee
which represents the DM flux in the direction of the galactic center
averaged over the typical ACT aperture of $\Delta \Omega = .001$ sr.
Extreme distributions, like the spiked profile \cite{27} and core
profile \cite{28}, correspond to increase and decrease of this flux
respectively by a factor of $\sim 10^3$.  

\begin{figure}[h!]
\begin{center}
\includegraphics[width=8cm]{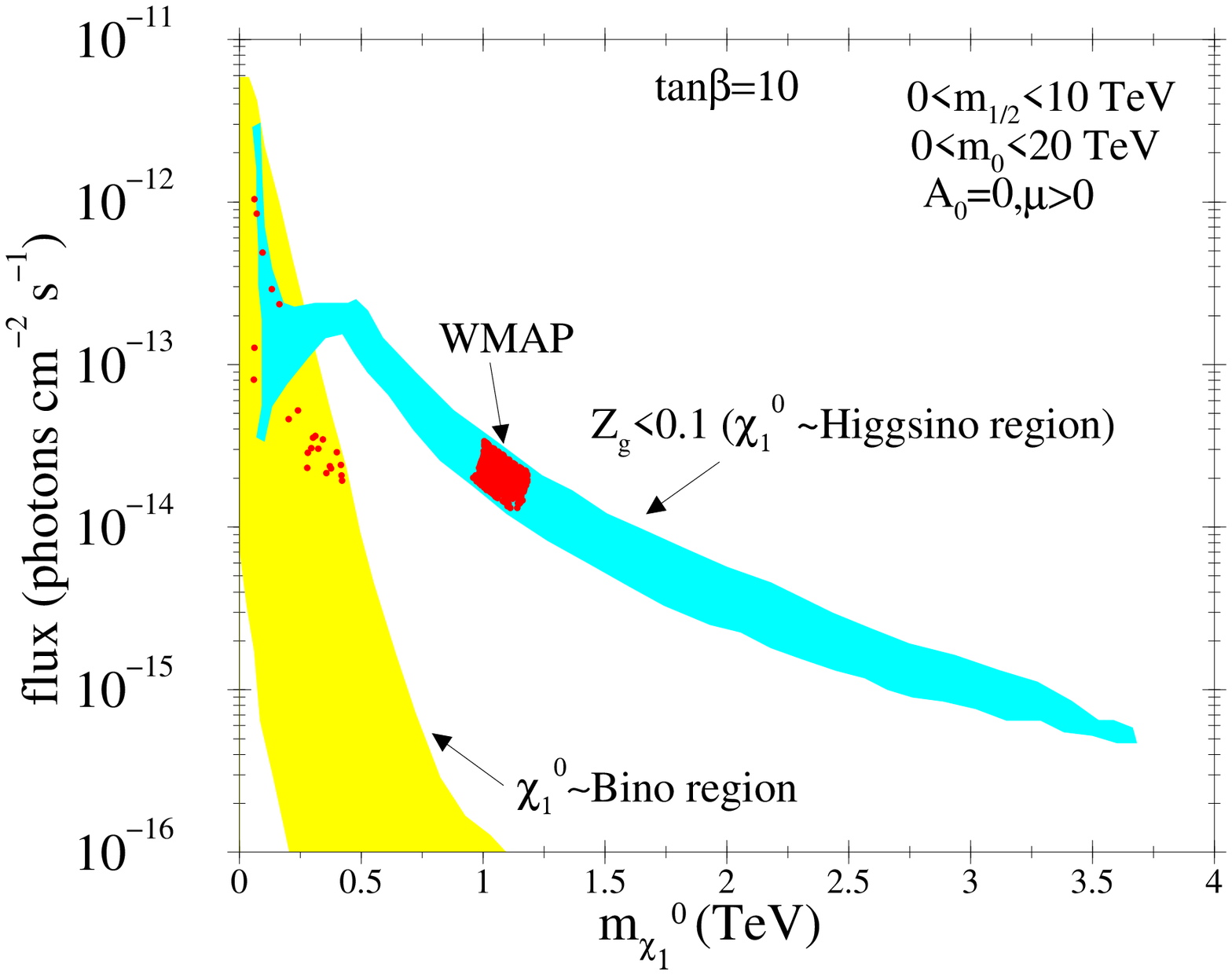}
\includegraphics[width=8cm]{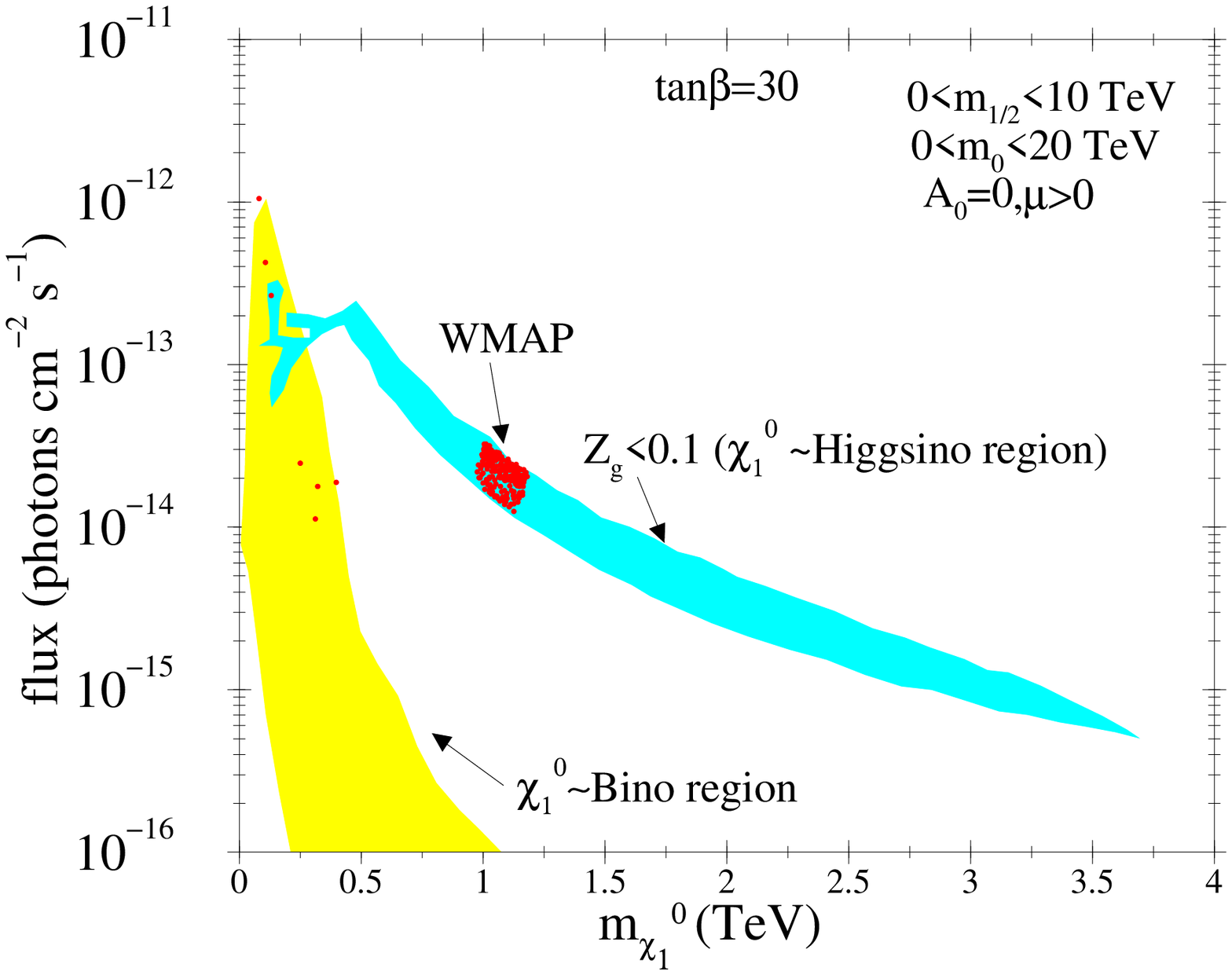}
\end{center}
\caption{Monochromatic gamma ray flux from DM pair annihilation near the
  galactic center shown for the NFW profile of DM halo distribution and an
  aperture $\Delta\Omega = 10^{-3} \ {\rm sr}$.}
\end{figure}

We have computed the $\gamma-$ray line signal (\ref{thirty}) for the NFW
profile and the aperture $\Delta\Omega = .001$ sr using the Dark SUSY code
\cite{29}. Fig.~5 shows the resulting signal against the DM mass, where we
have added the $\gamma\gamma$ and $\gamma Z$ contributions, since they give
identical photon energy $(= m_\chi)$ within the experimental resolution. This
result agrees well with that of ref. \cite{30}. The vertical spread in the
higgsino band reflects the dependence of the annihilation cross section
(\ref{twentyeight}) on the mass difference $\delta m_c$. As noted in \cite{30}
the discovery limit of the above-mentioned ACT experiments goes down to
$10^{-14} \ {\rm cm}^{-2} {\rm s}^{-1}$. Thus for the NFW profile one expects
a $\gamma-$ray line signal that should be detectable for the WMAP favored mass
range of $m_\chi \simeq 1$ TeV. Recall that the signal rate will go up by a
factor or of $\sim 10^3$ for the spiked profile \cite{27}, while it will go
down by a similar factor for the core profile \cite{28}. In the latter case it
will fall below the discovery limit of these ACT experiments.

In fact, H.E.S.S. {\em did} detect TeV photons coming from the direction of
the galactic center \cite{hess}. However, they observe a continuous spectrum
extending beyond 10 TeV in energy, which can be described quite well by a
power law. This is not what one expects from DM annihilation.  In fact, the
spectrum looks very similar to that of other ``cosmic accelerators'' observed
by the H.E.S.S. telescopes. It is currently not clear whether this signals
comes right from the center of our galaxy (defined as the location of a
supermassive black hole), or from a nearby supernova remnant (SNR); in
particular the SNR Sagittarius A East might be the culprit \cite{gm}. Note
that SNR are known to emit TeV photons with power--law spectra. The H.E.S.S.
collaboration is now working on improving their angular resolution. If the
source of the observed TeV photons, whose flux is well above the detectable
limit, continues to coincide with the galactic center within the resolution,
the discovery of a line signal from DM annihilation at the center would become
more difficult, since one would then have to look for a peak in the spectrum
on top of a sizable smooth background.

\section*{5. Higgsino LSP in other SUSY Models}

We have so far concentrated in the mSUGRA model for its simplicity and economy
of parameters. However one can easily see that all our results hold for the
higgsino LSP in a host of other SUSY models. This is because the relevant
interactions are the higgsino interaction with the gauge bosons, which are
completely determined by the gauge charges of $\tilde H_1$ and $\tilde H_2$
along with their mixing. Both these features are common to all variants of the
minimal supersymmetric standard model (MSSM). Thus the cross sections for the
higgsino annihilation processes (\ref{eleven}), and the resulting higgsino
masses (\ref{ten}) obtained using the constraint (\ref{eight}) on their relic
density, are common to all MSSMs, as long as the other superparticles and
heavy Higgs bosons have masses $\gsim 2$ TeV. The same is true for the signals
depicted in Figs.~2--5. This also explains why the DM results of Figs. 1 and 5
are essentially independent of all SUSY parameters except for the higgsino
masses.  We therefore have made no mention of these parameters while
presenting the collider signals of Figs.~2--4. It should be mentioned here
that the strip to the left of the DM allowed higgsino LSP range in Fig.~1
corresponds to an underabundance of DM relic density in the standard
cosmological model.  However, additional thermal and nonthermal mechanisms of
DM production have been suggested \cite{31,32}, which could enhance the DM
relic density over its standard cosmological model value. In the presence of
such mechanisms the DM allowed range will move to the $\mu < 1$ TeV region;
and so will the higgsino LSP mass. In that case the collider and DM signals
shown in Figs. 2,3 and 5 over the LSP mass range of 200 -- 1000 GeV will
become relevant.

We saw above that in the context of mSUGRA, a TeV higgsino can be the LSP only
if sfermions lie near 10 TeV or even higher. This is adequate for suppressing
FCNC processes even without assuming flavor universality of scalar soft
breaking parameters. It also allows ${\cal O}(1)$ phases in the soft breaking
sector without violating constraints on CP violating processes. This greatly
opens up the allowed parameter space, even if one keeps the two soft breaking
Higgs masses the same in order to achieve radiative EWSB. Examples for such
models are the so--called inverted hierarchy and more minimal supersymmetry
models \cite{15}. At the cost of additional finetuning(s) \cite{16}, one can
even entertain the idea of moving the sfermion masses to yet larger values, as
in the split SUSY model \cite{4}.

We saw in Figs.~1 that mSUGRA predicts the LSP to be Bino--like over most of
the theoretically allowed parameter space. This can be traced back to the fact
that the coefficient $C_2$ in eq.(\ref{five}) is quite large and positive,
while $|C_1|$ is small, so that $|\mu| > M_1$ at the weak scale unless $m_0^2
\gg m_{1/2}^2$. $C_1$ can be increased if the Higgs soft breaking masses
exceed the stop masses at the GUT scale \cite{33}. On the other hand, $C_2$
can be reduced if the ratios $M_1/M_3$ and/or $M_2/M_3$ are increased
\cite{34} relative to their mSUGRA values (\ref{three}). Models with
non--universal scalar and/or non--universal gaugino masses therefore often can
accommodate a higgsino--like LSP more easily than mSUGRA does. Finally, if one
reduces the input scale, i.e. the scale where supersymmetry breaking is
mediated to the visible sector \cite{spaniards}, one simultaneously increases
$C_1$ and reduces $C_2$, again making it easier to obtain a higgsino--like
LSP. All our results will apply equally to these models.

\section*{6. Summary and Conclusions}

We have seen that a higgsino--like LSP can be Dark Matter in a variety of
supersymmetric models. In the most constrained case, the mSUGRA model, this
remains a possibility for arbitrarily large values of $m_0$ and $m_{1/2}$,
thereby greatly enlarging the cosmologically allowed region of parameter
space. As discussed in Sec.~5. a higgsino--like LSP can also be realized in
many extensions of the mSUGRA model.

In standard cosmology, and assuming that the LSP was in thermal equilibrium
after the period of last entropy production, the LSP relic density can be
calculated uniquely from its (co--)annihilation cross sections. In case of
higgsino--like LSP one finds that a mass near 1 TeV is required. This makes
sparticle searches at colliders quite challenging. In most models strongly
interacting sparticles have masses at least a factor of 5 above the LSP mass;
this is true in particular for all models with (approximate) gaugino mass
unification near the scale of Grand Unification. This means that the usual
SUSY signatures at the LHC will not work. We found in Sec.~3 that the
production of two higgsino--like states in vector boson fusion also does not
give rise to a detectable signal at the LHC if these states lie near 1 TeV.
Moreover, the energy of the next (international) linear collider ILC will not
be sufficient to produce pairs of TeV sparticles.

We therefore have to consider more futuristic colliders. We saw in Sec.~3 that
the proposed 3 TeV $e^+e^-$ collider CLIC offers quite good prospects, {\em
  if} the level of beamstrahlung induced underlying events can be kept under
control. This can be achieved by designing the accelerator such that the flux
of beamstrahlung photons remains small, and/or by building a sufficiently
sophisticated detector so that the kinematic distributions of soft particles
produced in two--photons events can be distinguished from those of the soft
decay products of the heavier higgsino--like states $\tilde \chi_2^0$ and
$\tilde \chi_1^\pm$. We also saw that the ability to polarize the incident
$e^\pm$ beams would be very helpful.

In order to show that a given particle forms the Dark Matter in the universe,
one will eventually have to detect these relics. We saw in Sec.~4 that in case
of a higgsino--like LSP the most promising search is that for a $\gamma-$ray
line at $E_\gamma \simeq m_\chi$. The flux of such photons should peak in
directions where DM particles accumulate. The by far most promising site is
therefore the center of our galaxy. Unfortunately here the signal might be
masked by the recently observed flux of TeV photons with a continuous spectrum
extending beyond 10 TeV. Improved angular and/or energy resolution would be
helpful in enhancing the signal to background ratio in this case.

We conclude that a TeV higgsino is a viable supersymmetric Dark Matter
candidate. The large sparticle masses characteristic of such a scenario
require some amount of finetuning, but alleviate problems with
flavor--changing neutral currents and CP violation. Testing this scenario
experimentally is challenging, but should be possible at future multi--TeV
$e^+e^-$ colliders like CLIC, and perhaps through the observation of a TeV
$\gamma-$ray line in atmospheric Cerenkov telescopes. Finding TeV higgsinos
either at colliders or in Dark Matter search experiments is certainly easier
than finding gravitinos, which have been much discussed lately as possible
Dark Matter candidates \cite{gravitino}. This scenario should therefore be
taken seriously, in particular if the LHC fails to discover supersymmetry.

\subsection*{Acknowledgments}
DC thanks the Physikalisches Institut of Bonn University for hospitality, and
DST, India for financial assistance under a Swarnajayanti Fellowship. The work
of MD was partially supported by the European Network for Theoretical
Astroparticle Physics, ENTApP. DPR and UC thank the organizers of the 8th
Workshop on High Energy Physics Phenomenology (WHEPP 8), where this
investigation was initiated.


\begin{thebibliography}{99}


\bibitem{1}
See e.g., Perspectives in Supersymmetry, ed. G.L. Kane,
World Scientific (1998); Theory and Phenomenology of Sparticles:
M. Drees, R.M. Godbole and P. Roy, World Scientific (2005).

\bibitem{2} 
Review of Particle Properties: S. Edelman et al,
Phys. Lett. {\bf B592}, 1 (2004).

\bibitem{3} 
See e.g., Higgs Boson Theory and Phenomenology: M. Carena
and H.E. Haber, Prog. Part. Nucl. Phys. {\bf 50}, 63 (2003).

\bibitem{4} 
N. Arkani-Hamed and S. Dimopoulos, JHEP {\bf 0506}, 073 (2005)
[hep--th/0405159]; 
G.F. Giudice and A. Romanino, Nucl. Phys. {\bf B699}, 65 (2004)
[hep--ph/0406088].

\bibitem{5} 
A.H. Chamseddine, R. Arnowitt and P. Nath, Phys. Rev. Lett. {\bf 49}, 970
(1982); R. Barbieri, S. Ferrara and C.A. Savoy, Phys. Lett. {\bf B119}, 343
(1982); L. Hall, J. Lykken and S. Weinberg, Phys. Rev. {\bf D27}, 2359 (1983).

\bibitem{6} 
L. Ib\'a\~nez and G.G. Ross, Phys. Lett. {\bf B110}, 215 (1982);
K. Inoue et al., Prog. Theor. Phys. {\bf 68}, 927 (1982).

\bibitem{7} 
M. Carena, M. Olechowski, S. Pokorski and C.E.M. Wagner,
Nucl. Phys. {\bf B426}, 269 (1994).

\bibitem{8} 
K.L. Chan, U. Chattopadhyay and P. Nath, Phys. Rev. {\bf D58}, 096004 (1998);
U. Chattopadhyay, A. Corsetti and P. Nath, Phys. Rev. {\bf D68}, 035005 (2003).

\bibitem{9} 
G. B\'elanger, F. Boudjema, A. Pukhov and A. Semenov, hep--ph/0405253.

\bibitem{10} 
C.L. Bennett et. al., Astrophys. J. Suppl. {\bf 148}, 1 (2003)
[astro--ph/0302207]; D.N. Spergel et al., Astrophys. J. Suppl. {\bf 148}, 175
(2003) [astro--ph/0302209] .

\bibitem{11} 
See e.g., J.R. Ellis et al., Nucl. Phys. {\bf B652}, 259
(2003); A. Bottino, F. Donato, N. Fornengo and S. Scopel, Phys. Rev. {\bf
  D68}, 043506 (2003) [hep--ph/0304080].

\bibitem{hpole}
A. Djouadi, M. Drees and J.--L. Kneur, hep--ph/0504090.

\bibitem{12} 
J.L. Feng, K.T. Matchev and T. Moroi, Phys. Rev. {\bf D61},
075005 (2000), and Phys. Rev. Lett. {\bf 84}, 2322 (2000); J.L. Feng,
K.T. Matchev and F. Wilczek, Phys. Rev. {\bf D63}, 045024 (2001).

\bibitem{13} 
U. Chattopadhyay, A. Datta, A. Datta, A. Datta and D.P. Roy, Phys. Lett. {\bf
  B493}, 127 (2000);
P.G. Mercadante, J.K. Mizukoshi and X. Tata, hep--ph/0506142;
H. Baer, T. Krupovnickas, S. Profumo and P. Ullio, hep--ph/0507282.

\bibitem{newtop}
The Tevatron Electroweak Working Group, hep--ex/0507091.

\bibitem{14}
J. Edsj\"o and P. Gondolo, Phys. Rev. {\bf D56}, 1879 (1997).

\bibitem{17} 
M. Drees, M.M. Nojiri, D.P. Roy and Y. Yamada, Phys. Rev. {\bf D56}, 276
(1997). 

\bibitem{15} 
A.G. Cohen, D.B. Kaplan and A.E. Nelson, Phys. Lett. {\bf B388}, 588 (1996).

\bibitem{18} 
A. Datta, P. Konar and B. Mukhopadhyaya, Phys. Rev. Lett. {\bf 88}, 181802
(2002). 

\bibitem{19} 
O. Eboli and D. Zeppenfeld, Phys. Lett. {\bf B495}, 147 (2000).

\bibitem{20} 
Physics at the CLIC Multi-TeV Linear Collider: CLIC Physics 
Working Group, hep--ph/0412251.

\bibitem{21} 
C.H. Chen, M. Drees and J.F. Gunion, Phys. Rev. Lett. {\bf 76},
2002 (1996) [hep--ph/9512230]; Phys. Rev. D55, 330 (1997)
[hep--ph/9607421]; Addendum/Erratum [hep--ph/9902309].

\bibitem{22} 
OPAL Collab., G.A. Abbiendi et al., Eur. Phys. J. {\bf C29}, 479 (2003).

\bibitem{dg1}
M. Drees and R.M. Godbole, Z. Phys. {\bf C59}, 591 (1993) [hep--ph/9203219].

\bibitem{dg}
M. Drees and R.M. Godbole, Phys. Rev. Lett. {\bf 67}, 1189 (1991).

\bibitem{23} 
See e.g., Physics Interplay of the LHC and the ILC: LHC/LC study group,
hep--ph/0410364. 

\bibitem{24} 
L. Bergstrom, P. Ullio and J.H. Buckley, Astropart Phys. {\bf 9}, 137 (1998)
[astro--ph/9712318]. 

\bibitem{25} 
P. Ullio and L. Bergstrom, Phys. Rev. {\bf D57}, 1962 (1998) [hep--ph/9706232].

\bibitem{newan}
L. Bergstrom, T. Bringmann, M. Eriksson and M. Gustafsson, hep--ph/0507229.

\bibitem{26} 
J.F. Navarro, C.S. Frenk and S.D.M. White, Astrophys. J. {\bf 462}, 563 (1996)
and {\bf 490}, 493 (1997). 

\bibitem{27} 
J. Diemand, B. Moore and J. Stadel, Mon. Not. Roy. Astron. Soc. {\bf 353}, 624
(2004); B. Moore et al., Astrophys. J. {\bf 524}, L19 (1999).

\bibitem{28} A. Burkert, Astrophys. J. {\bf 447}, L25 (1995).

\bibitem{29} 
P. Gondolo, J. Edsj\"o, P. Ullio, J. Bergstrom, M. Schelke and E.A. Baltz,
JCAP {\bf 0407}, 008 (2004) [astro--ph/0406204]. 

\bibitem{30} 
J. Hisano, S. Matsumoto, M.M. Nojiri and O. Saito, hep--ph/0412403; J. Hisano,
S. Matsumoto and M.M. Nojiri, Phys. Rev. {\bf D67}, 075014 (2004), and
Phys. Rev. Lett. {\bf 92}, 031303 (2004). 

\bibitem{hess}
The H.E.S.S. Collab., F. Aharonian et al., astro--ph/0408145.

\bibitem{gm}
D. Grasso and L. Maccione, astro--ph/0504323.

\bibitem{31} 
P. Salati, Phys. Lett. {\bf B571}, 121 (2003) [astro--ph/0207396]; 
F. Rosati, Phys. Lett. {\bf B570}, 5 (2003) [hep--ph/0302159].

\bibitem{32} 
B. Murakami and J.D. Wells, Phys. Rev. {\bf D64}, 015001
(2001); T. Moroi and L. Randall, Nucl. Phys. {\bf B570}, 455 (2000);
M. Fujii and K. Hamaguchi, Phys. Lett. {\bf B525}, 143 (2002); W.B. Lin,
D.H. Huang, X. Zhang and R.H. Brandenburger, Phys. Rev. Lett. {\bf 86},
954 (2001).

\bibitem{16} M. Drees, hep--ph/0501106.

\bibitem{33} 
J.R. Ellis, T. Falk, K.A. Olive and Y. Santoso, Nucl. Phys. {\bf B652}, 259
(2003). 

\bibitem{34} 
U. Chattopadhyay and D.P. Roy, Phys. Rev. {\bf D68}, 033010 (2003).

\bibitem{spaniards}
E. Gabrielli, S. Khalil, C. Mu\~noz and E. Torrente--Lujan,
Phys. Rev. {\bf D63}, 025008 (2001) [hep--ph/0006266].

\bibitem{gravitino}
See e.g. J.R. Ellis, K.A. Olive, Y. Santoso and V.C. Spanos,
Phys. Lett. {\bf B588}, 7 (2004), [hep--ph/0312262];
J.L. Feng, S. Su and F. Takayama, Phys. Rev. {\bf D70}, 075019 (2004)
[hep--ph/0404231];
L. Roszkowski and R. Ruiz de Austri, hep--ph/0408227.

\end{thebibliography}
\end{document}